\documentclass[useAMS,usenatbib]{mn2e}

\usepackage{times}
\usepackage{graphics}
\usepackage{amsmath}
\usepackage{amssymb}
\usepackage{listings}
\usepackage{array}
\usepackage{bm}
\usepackage{graphicx}
\newcommand{\bd}{\begin{displaymath}}
\newcommand{\ed}{\end{displaymath}}

\newcommand{\be}{\begin{equation}}
\newcommand{\ee}{\end{equation}}
\newcommand{\bea}{\begin{eqnarray}}
\newcommand{\eea}{\end{eqnarray}}

\title[Black Holes in Globular Clusters]
{Mergers and ejections of black holes in globular clusters}
\author[Aarseth]{
	Sverre J.~Aarseth\thanks{ E-mail:sverre@ast.cam.ac.uk } \\
Institute of Astronomy, University of Cambridge, Madingley Road,
Cambridge, CB3 0HA, UK }
\begin{document}
\date{Accepted 2012 January 31. Received 2012 January 31; in original form 2011 November 11}

\pagerange{\pageref{firstpage}--\pageref{lastpage}} \pubyear{2011}

\maketitle

\label{firstpage}

\begin{abstract}
We report on results of fully consistent $N$-body simulations of globular
cluster models with $N$ = 100~000 members containing neutron stars and
black holes.
Using the improved `algorithmic regularization' method of Hellstr\"om and Mikkola
for compact subsystems, the new code {\tt NBODY7} enables for the first time
general relativistic coalescence to be achieved for post-Newtonian terms
and realistic parameters.
Following an early stage of mass segregation, a few black holes form a
small dense core which usually leads to the formation of one dominant binary.
The subsequent evolution by dynamical shrinkage involves the competing
processes of ejection and mergers by radiation energy loss.
Unless the binary is ejected, long-lived triple systems often exhibit
Kozai cycles with extremely high inner eccentricity ($e > 0.999$) which may
terminate in coalescence at a few Schwarzschild radii.
A characteristic feature is that ordinary stars as well as black holes and
even BH binaries are ejected with high velocities.
On the basis of the models studied so far, the results suggest a limited
growth of a few remaining stellar mass black holes in globular clusters.
\end{abstract}


\section{INTRODUCTION}

Recent years have seen many studies relating to the dynamics of black holes (BHs)
both in galactic and extra-galactic systems.
In view of the observations of the S stars at the galactic centre it is not
surprising that most efforts have been directed towards a relatively massive BH.
However, there is also considerable interest in the effect of BHs in star clusters.
It has been argued that because velocity kicks may occur at formation, some BHs
are more likely to be retained in globular rather than open clusters.
We may therefore distinguish between problems dealing with several stellar mass
BHs or one dominant body formed by an accretion process.
As far as $N$-body simulations are concerned, these two types call for different
methods of solution.
Thus in the former we need to treat strong interactions of BHs, while in the
latter case a number of short binary periods require careful attention.
A closer inspection of investigations concerned with a relatively massive BH
shows that Newtonian motions are often adopted in full simulations
(Brockamp, Baumgardt \& Kroupa 2011).
On the other hand, investigations involving binary BHs tend to avoid the
numerical problems associated with small Schwarzschild radii by artificial
scaling of the masses or magnifying the post-Newtonian effects
(Iwasawa, Funato \& Makino 2006, Berentzen et al. 2009).

The first full N-body simulation with post-Newtonian (PN) terms (Aarseth 2003b)
indicated that a black hole binary of intermediate mass may achieve the general
relativistic (GR) coalescence condition following eccentricity growth by the
Kozai mechanism (Kozai 1962, 2004).
This development was made possible by a special integration method for a
long-lived massive binary (Mikkola \& Aarseth 2002, {\tt TTL}).
An alternative method, suitable for a dominant single BH, showed several
examples of GR coalescence for realistic parameters (Aarseth 2007).
Although expensive, the so-called wheel-spoke method generalized from a
three-body regularization algorithm (Aarseth \& Zare 1974; Zare 1974) is
accurate and enables extremely close orbits to be studied.
It is notable that in both these investigations, Kozai cycles played an
important role, with maximum eccentricities often reaching large values
(e.g.~0.9999)\footnote{Early $N$-body simulations report large eccentricity
growth to $e > 0.999$ at constant semi-major axis in triple systems which
exhibit the hallmarks of Kozai cycles (van Albada 1968, Aarseth 1971).}.

The motivation for the present investigation goes back more than 40 years.
Thus an early simulation demonstrated that two massive bodies (factor of 5)
gave up kinetic energy to the other members and reached the centre after
a few crossing times (Aarseth 1971).
Since most of the other core particles were expelled, the two dominant
members invariably formed a binary.
Although the cluster membership was only $N = 250$, the mass segregation
still operated in a qualitatively similar way.
Given a dynamically shrinking massive binary, the probability of a long-lived
triple with suitable inclination is non-negligible, especially since any
stellar mass suffices for inducing Kozai cycles.

Further simulations of small globular clusters with $N = 10^5$ members
containing a significant BH component were made with special-purpose GRAPE
computers (Mackey et al.~2008).
More recently similar systems have been studied with Graphics Processing
Units (GPU) which also require special programming (Nitadori 2009).
The software implementation in the standard {\tt NBODY6} code based on the
fourth-order Hermite block-step neighbour scheme has been described in a recent
paper (Nitadori \& Aarseth 2012).
Of special interest here is an application to compact subsystems of more than
two dominant members, all of which may be BHs.
In particular, we wish to explore possible post-Newtonian dynamics in a globular
cluster context.
Although there have been attempts of including the 2.5PN radiation term using
the standard {\tt NBODY6} code (Aarseth 2003a) by others
(Banerjee, Baumgardt \& Kroupa 2010), simplified treatments fall short of
realistic requirements and hence a fresh approach is called for.

In this paper, we make use of a new powerful integration package that can overcome
the main numerical difficulties associated with compact subsystems.
Here the challenge is to treat stellar mass BH binaries of extremely short
periods during late stages of inspiral using realistic parameters.
The improved algorithmic regularization chain method
(Hellstr\"om \& Mikkola 2010, {\tt ARC})
traces its development from chain regularization (Mikkola \& Aarseth 1993, {\tt CHAIN})
via the logarithmic Hamiltonian method (Mikkola \& Tanikawa 1999) and the
time-transformed leapfrog code {\tt TTL} to special treatments of post-Newtonian
terms (Mikkola \& Merritt 2008), also denoted algorithmic regularization.
The resulting code called {\tt NBODY7} is able to deal with extreme few-body
configurations up to the final stages of GR coalescence without recourse to
artificial scaling of the parameters.
A key feature of the present investigation is that velocity kicks are assigned to 
neutron stars and BHs at formation, such that only about 10 percent of the membership
is retained in both cases.
The subsequent evolution by mass segregation of the heaviest members gives rise to
compact subsystems which sometimes require post-Newtonian treatments due to the process
of Kozai eccentricity cycles.
Consequently, even a small population of stellar mass black holes are able to
control the central region.
The challenge is then to employ an efficient method for dealing with extremely
compact subsystems where PN terms may be needed for the dominant two-body motion.

This paper is organized as follows.
We begin by outlining the $N$-body implementation of algorithmic regularization and
post-Newtonian terms for a compact subsystem.
Section 3 is devoted to a general description of mass segregation while results
for particular models are illustrated in section 4.
Finally, some aspects of the simulations are discussed in section 5, followed by
conclusions.

\section{COMPACT SUBSYSTEMS}

Simple test calculations with a few free-floating stellar mass BHs added reveal
a characteristic behaviour of mass segregation in accordance with expectations.
The essential question is concerned with the degree of central concentration;
namely when does the innermost core stop shrinking due to binary activity.
Some preliminary investigations were sufficiently encouraging to proceed with
new code developments as summarized below.

The main challenge for dealing with a compact configuration is to maintain
accuracy, albeit at increased cost.
This involves some kind of regularization to avoid near-singular force terms.
The adopted method (Mikkola \& Merritt 2006, 2008) is based on a sophisticated
time transformation that enables even collision orbits to be integrated at high
accuracy.
Briefly, the underlying integrator (Bulirsch \& Stoer 1966) senses an approach
to collision and avoids evaluating the force at singular points.
Moreover, using a judicious choice of coefficients in the time transformation,
the problem of large mass ratios can also be accommodated.
Two analogous forms of the time transformations are used for a subsystem of
$N_{\rm ch}$ interacting particles (Mikkola \& Merritt 2008),
\be
ds = [\alpha (T + B) + \beta \omega + \gamma ] dt =
     [ \alpha U + \beta \Omega + \gamma ] dt,
\ee
where $ds$ is the new differential time element and $\alpha, \beta, \gamma$
are adjustable constants.
Furthermore, $T$ is the kinetic energy, $U$ the (positive) potential energy,
$B$ the binding energy, $B = U - T$, and $\Omega$ is an optional function of
the coordinates. 
It can be shown that this relation provides well-behaved solutions for two-body
collisions when used in connection with a simple leapfrog algorithm where the
results are improved by an extrapolation method.
Moreover, the case of velocity-dependent perturbations (e.g. post-Newtonian
terms) can be treated explicitly with high accuracy (Hellstr\"om \& Mikkola 2010).

The alternative time transformations are applied to coordinate and velocity
leapfrog integrations, respectively.
Thus the $\Omega$-formulation is related to the {\tt TTL} method in which 
$\dot \omega = \pmb {v}{\cdot} \nabla \Omega $ provides a
regular solution for $\omega(t) = \Omega(t)$ (Mikkola \& Aarseth 2002).
When a range of masses are involved, the choice
$(\alpha, \beta, \gamma) = (1.0, 0.001, 0)$ is recommended, where the non-zero
value of $\beta$ yields increased accuracy for any massive bodies.
Note that the algorithmic chain regularization actually employs the chain data
structure without solving the corresponding equations of motion.
This procedure leads to significant reduction of round-off errors and therefore
plays a key role in the formulation.

The choice of the subsystem membership is to some extent experimental.
Thus we must consider the cost of treating a small subsystem by
an accurate method where a larger size would also necessitate a greater
number of perturbers.
Note that the basic integrator performs a large number ($\sim 100$) of
function evaluations per step and hence also requires many coordinate
predictions of the perturbers for consistency.
Although very generous choices of the perturber number have been used
before (Harfst et al. 2008), it has yet to be demonstrated that this is
necessary.
A careful investigation of the corresponding energy change reveals that the
external effect is relatively small for dimensionless perturbations beyond
$\gamma_{\rm pert} \sim 10^{-9}$ which is a typical limit for selection
(reduced from the earlier value $10^{-7}$).
Even so, perturber numbers of only 3 or 4 are typical in the present simulations
with small chain memberships\footnote{This behaviour is characteristic for a
dominant binary.}.
More problematic are the decisions about when a perturber should be included
in the membership and vice versa.

The software package {\tt ARC} replaces the procedures relating to the
{\tt CHAIN} code, with analogous decision-making for the interface connection
to the standard {\tt NBODY6}.
Consequently, the new code is called {\tt NBODY7}.
Although {\tt ARC} contains post-Newtonian terms up to 3.5PN, an alternative
formulation up to 3PN (Blanchet \& Iyer 2003, Mora \& Will 2004) has been
retained (Aarseth 2007).
This has the advantage that relativistic expressions for the semi-major axis and
eccentricity are readily available for decision-making and data analysis.

We now review some relevant aspects relating to the implementation of PN terms.
The equation of dominant two-body motion with $G = 1$ is written as
\be
\frac {d^2 {\bf r}}{d t^2} =
 \frac {M}{r^2} \left[(-1 + A) \frac {{\bf r}}{r} + B {\bf v} \right ],
\ee
where $M = m_1 + m_2$ and $A$ and $B$ represent the post-Newtonian terms for
coordinates and velocities, $\bf r$ and $\bf v$, respectively.
The perturbing {\it force} then takes the form
\bea 
 {\bf P}_{\rm GR} \,&=&\,
\frac {m_1 m_2}{c^2 r^2} \biggl[ (A_1 + \frac {A_2}{c^2} +
 \frac {A_{5/2}}{c^3} + \frac {A_3}{c^4}) \frac {\bf {r}}{r}  + \nonumber \\
 && ~~~~~~~~~~~~~~~
(B_1 + \frac {B_2}{c^2} + \frac {B_{5/2}}{c^3} + \frac {B_3}{c^4}) \bf{v} \biggr] .
\label{eq:GR}
\eea
In $N$-body applications, the scaled speed of light is formally given by
$c = 3 \times 10^5/ V^{\ast}$, with $V^{\ast}$ the velocity unit
(typically $16~{\rm km~s^{-1}}$ here).
However, much smaller values are often used by other workers for easing numerical problems.
The expansion coefficients in $A, B$ to sixth order in $1/c$ are obtained from
the current formulation.
Although the spin term is of order $c^{-3}$, it has only been used in tests here
for the most dominant body (Gopakumar, private communication).
One reason for this neglect is that the spin effect
depends on the uncertain initial magnitude as well as direction.
However, only the radiation term gives rise to energy dissipation.

The time-scale for gravitational radiation (in $N$-body units) is given by the
classical expression (Peters 1964)
\be \tau_{{\rm GR}} = \frac {5 }{64} \frac {a^4 c^5}{X (1+X) m_1^3 }
  \frac {(1 - e^2)^{7/2}} {g(e)} \,,
 \label{eq:TGR}
\ee
where $X = m_2/m_1$ with $m_1 > m_2$ and $g(e)$ a known function of eccentricity
($\simeq 4.5$ for large $e$).
In the relativistic regime, the binding energy per unit mass is determined from
\be
\epsilon_b = \epsilon_0 + \frac {\epsilon_1}{c^2} +
\frac {\epsilon_2}{c^4} + \frac {\epsilon_3}{c^6},
\ee
where $\epsilon_0$ is the Newtonian value. This yields the semi-major axis
\be
a = -\frac {M}{2 \epsilon_b} .
\ee
Likewise, the eccentricity is obtained via the angular momentum expansion
\be
{\bf J} = {\bf J}_0 (1 + \frac {f_1}{c^2} + \frac {f_2}{c^4} )
\ee
by
\be
 e^2 = (1 - \frac {{\bf J}^2} {M a}) \,.
\ee
The coefficients $\epsilon_1, \epsilon_2, \epsilon_3, f_1, f_2$ are also taken
from the literature cited above.

Another useful quantity is the Lenz vector
\be
 {\bf e} = \frac { {\bf v} \times {\bf r} \times {\bf v} } {M} - \frac {{\bf r}} {r}
\ee
which can be used to determine the Einstein (1915) pericentre shift
(in radians per orbit)
\be
 \Delta w \,=\, \frac {6 \pi M} {c^2 a (1 - e^2) } \,.
 \label{eq:dw}
\ee

An effort has been made to maintain a high level of energy conservation.
For this purpose we split the contributions to subsystem energy changes
into two parts.
Thus the integrated effect of particle perturbations is used to monitor the
instantaneous internal energy while additional contributions arising from the
PN terms are treated separately.
Although only the gravitational radiation is dissipative, the other terms
depend on the orbital parameters and therefore need to be included for
conservation purposes.
Consequently the sum of the internal energy and the accumulated relativistic
energy change is added to the standard $N$-body value based on considering the
subsystem as a point-mass body.

The question of when to include PN terms is a delicate one.
In view of the increasing cost of the higher orders, we have devised a
scheme based on efficiency which has proved itself (Aarseth 2007).
The basic idea is to delay the PN stage until the radiation time-scale falls
below a specified value; e.g. $\tau_{\rm GR} < 500$ or about 30~Myr.
At this stage, the first-order precession terms of comparable cost are also
activated, especially because a high precession rate may affect any
eccentricity growth.
The second and third-order terms are then included once the time-scale shrinks
to 50 and 1 $N$-body units, respectively.
Inspection of actual examples show that once orbital shrinkage begins, a
significant acceleration often takes place, boosted by favourable Kozai cycles.

As an additional safeguard, the next order is also activated if the Kozai
period (Kiseleva, Eggleton \& Mikkola 1998), $T_{\rm Kozai}$, falls below
a certain value.
This time-scale can be evaluated for long-lived triple configurations which
are often present.
The ultimate aim of post-Newtonian $N$-body simulations is to see whether
coalescence can occur for realistic parameters; in this case stellar mass
BHs in globular clusters.
Although even smaller two-body separations can be reached with high accuracy,
we now define coalescence at four Schwarzschild radii by
\be
R_{\rm coal} = \frac {8 M} {c^2}.
\label{eq:Rcoal}
\ee
As can be seen from equation~(\ref{eq:TGR}), the corresponding time-scale 
becomes extremely small during the final stages\footnote{Typical parameters
are $a=1 \times 10^{-10}, c = 2 \times 10^4, m_1 = 3 \times 10^{-4}. $ }.

Termination by coalescence may also be defined at an earlier stage if
conditions are favourable.
We distinguish between three different cases; (i) $N_{\rm ch}=2$ with no
perturbers and small $\tau_{\rm GR}$ (e.g. $\tau_{\rm GR} < 1$),
(ii) $a (1 - e) < R_{\rm coal}$ during
the second or third PN stage, and (iii) $N_{\rm ch} = 3, T_{\rm Kozai} > 25$,
likewise for small $\tau_{\rm GR}$ and large outer pericentre (factor of 100).

At later times it frequently happens that a dominant binary may not have any
perturbers and the GR time-scale is large, in which case it is treated in the
usual unperturbed two-body approximation; i.e. with the internal motion frozen.
A new subsystem is then re-initialized (at the same orbital phase) once a
binary perturber is identified, based on several conservative procedures involving
relative distances and velocities (Aarseth 2003a, p.190) \footnote{Following
initialization, the relative perturbation is typically $< 10^{-7}.$}.

The {\tt ARC} system is advanced in a similar way to the standard {\tt CHAIN}
code.
Thus each block-step time interval is treated sequentially before control is
returned to the main code which also deals with regularized binaries in a
similar manner.
A full discussion of the GPU implementation in {\tt NBODY6} has been presented
elsewhere (Nitadori \& Aarseth 2012).

\section{MASS SEGREGATION}

We have performed a number of $N$-body simulations in order to improve the
statistical results of rare events.
Equilibrium Plummer models are used with $N = 1 \times 10^{5}$
particles and an IMF mostly in the range $50 - 0.1~M_{\odot}$ 
(Kroupa, Tout \& Gilmore 1993) extended into the brown dwarf regime.
We adopt standard $N$-body units with $G=1$ where the total mass and energy
are scaled to 1 and $E = -1/4$, respectively, with mean square velocity of 1/2.
Physical units are then readily obtained once the length scale ($R^{\ast}$ in pc)
and mean mass ($\bar m = 1/N$ in $M_{\odot}$) are specified.
For most of the present models $R^{\ast} = 1~$pc and $\bar m = 0.6~M_{\odot}$,
with a scaled half-mass radius $r_{\rm h} \simeq 0.8$ and time-scale
conversion factor $T^{\ast} = 0.06~$Myr.
Likewise, the typical velocity scale unit is $V^{\ast} \simeq 16~{\rm km~s^{-1}}$.
Since the speed of light is traditionally taken as a free parameter in such
simulations, we have adopted $c = 18\,000$ for the final models, close to the
actual value.

Before presenting results, it may be of interest to consider the justification
for the post-Newtonian implementation in the context of the present parameters.
Let us assume an energetic binary, formed by dynamical means with
semi-major axis $a_{\rm hard} \simeq 2/N = 2 \times 10^{-5}$ and component
masses $20~M_{\odot}$.
The energy of the massive binary would then represent 1~percent of
the total energy which can be achieved in this type of simulation.
Moreover, from previous experience (Aarseth 2007), a maximum eccentricity of
0.999 is often seen.
According to equation~(\ref{eq:TGR}), the corresponding time-scale would be
35 $N$-body units or 2~Myr which is practically feasible, although it would
require a very large number of binary orbits.
Moreover, starting the PN sequence at such high eccentricity usually implies
the presence of a Kozai cycle which would tend to counteract the decay and
hence assist in reducing the actual time interval further.

\begin{figure}
 \includegraphics[width=8.cm]{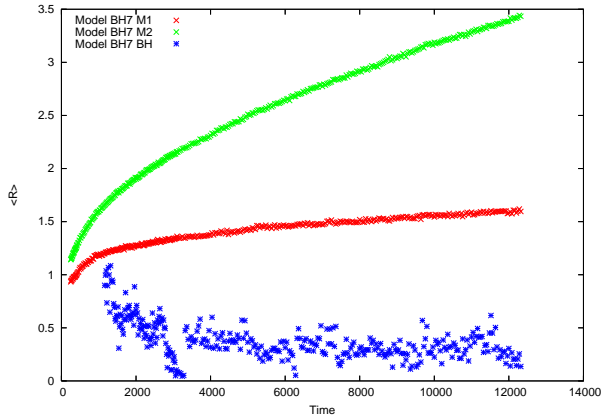} 
     \caption{~~Mean central distance for light masses (upper curve),
                 heavy masses (middle curve) and BHs (lower curve), Model BH7.
                 Because of mass-loss evolution, the maximum stellar mass has
                 been reduced to $3.1~M_{\odot}$ at the end (750~Myr), with
                 the mid-point and lower values at $0.61~M_{\odot}$ and
                 $0.10~M_{\odot}$, respectively.
                 A small number of BHs are included in the heavy population
                 but contribute reduced values. }
   \label{fig:mt}
\end{figure}

The full $N$-body simulations employ the {\tt NBODY6}  synthetic stellar evolution
package (Hurley, Pols \& Tout 2000, 2002) which provides information about stellar
mass, radius and type as a function of time.
Of particular relevance here are the final BH masses (Eldridge \& Tout 2004).
Thus for an intermediate metallicity of $Z = 0.001$ each model typically yields
about 800 neutron stars and 140 BHs.
The latter are in the range 3.0 to $19~M_{\odot}$ with an average mass of
$8~M_{\odot}$.
In the absence of a consensus on BH and neutron star velocity kicks at formation,
we adopt a conservative approach.
Thus the kick distribution is chosen from a Maxwellian with dispersion
$2 V^{\ast} \simeq 32~{\rm km~s^{-1}}$ and applied to both populations, which gives a
retention of about 10~percent.
As usual with tidal fields, escaping stars are removed at twice the tidal
radius, where $r_{\rm tide} \simeq 56~$pc.

The increased velocity dispersion for BHs due to velocity kicks is evident during
the time interval of $4 - 10$~Myr, followed by neutron star formation up to $60$~Myr.
A further stage of escape reduces the populations to about 14 and 40, respectively,
with an average BH mass of $9~M_{\odot}$ at age $100$~Myr.
By this time, the mass segregation of BHs is well developed.
A typical behaviour is exhibited in Fig.~1 which shows the mean central distance
of two populations, each containing 50~percent of the total mass for Model~BH7.
The third plot gives the similar information for the BHs.
Since the BH membership is relatively small, the geometrical mean separation
is used here.
This definition is more representative when the result may be affected by
a few outliers, including escapers.
Unless specified otherwise, all quantities plotted are in $N$-body units, where
the scaling factor for time in Myr is 0.06.

We note the characteristic behaviour that the mean central distance for the heavy
members also increases with time\footnote{The contrast is enhanced further using
the geometric mean separation.}.
The small BH population exhibits significant contraction subject to sporadic
expansion.
The rapid expansion of the core from small values ($\simeq 0.05$) near the times
3200 and 6300 is connected with significant binary activity.
In fact, the first BH binary is created at the former time
($a \simeq 5 \times 10^{-4}$).
Further shrinkage of the expanded core then takes place until a similar minimum
is reached whereupon an existing binary experiences exchange with fast ejection
($v_{\rm ej} \simeq 48~{\rm km~s^{-1}}$) of a massive ($11~M_{\odot}$)
component.
It is worth noting that already at $t \simeq 3000$ (or 180~Myr) only 14 BHs remain.
A third notable event occurs near $t \simeq 9600$ when the binary is ejected in
a slingshot interaction from a bound triple.
The relative centre of mass (c.m.) velocities of the ejected members,
20 and $63~{\rm km~s^{-1}}$, mirror closely the mass ratio of 3.2.
Thus we see four distinct stages of core collapse, with only six relatively
light BHs remaining when the calculations were terminated.

The GR radiation time-scales of ejected BH binaries are of potential interest for
future detectors.
In the absence of a systematic study, we note that nearly all the values
predicted by equation~(\ref{eq:TGR}) are well below 10~Gyr with 10-100~Myr being
typical, in qualitative agreement with provisional findings (Banerjee et al. 2010).

\begin{figure}
 \includegraphics[width=8.cm]{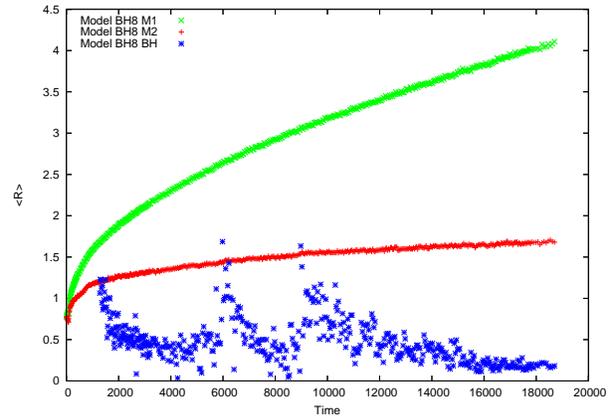} 
     \caption{~~Mean central distance for light masses (upper curve),
                 heavy masses (middle curve) and BHs (lower curve), Model BH8.
                 The mass ranges are similar to Fig.~1.}
   \label{fig:mt}
\end{figure}

More pronounced evidence of oscillatory core behaviour can be seen in Fig.~2
which displays the same quantities for Model~BH8.
Here a genuine GR coalescence occurs near $t \simeq 5060$ but this does not
induce core expansion.
However, significant binary activity occurs around $t \simeq 5950$ which
results in the fast ejection of a massive BH and subsequent expansion.
The next noticeable expansion is at $t \simeq 8970$ when a massive triple
escaped by the slingshot mechanism.
Again the large ejection velocities of 29 and $121~{\rm km~s^{-1}}$ with
respective masses 45 and $11~M_{\odot}$ confirm nicely the momentum
conservation.
It should be emphasized that the slingshot events above were purely Newtonian,
with the final transition from
$a \simeq 8 \times 10^{-6} ~{\rm to}~ 4 \times 10^{-6}$.

\begin{figure}
 \includegraphics[width=8.cm]{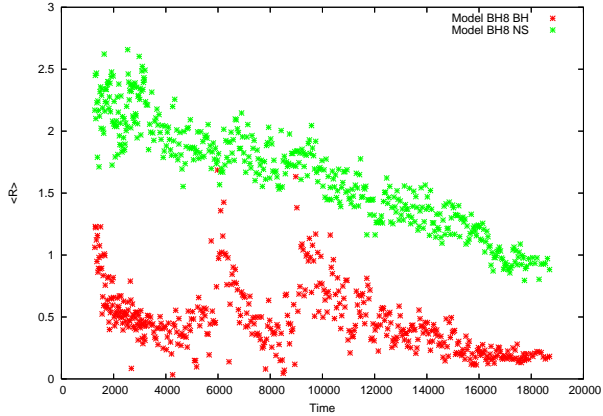} 
     \caption{~~Geometric mean central distance for neutron stars
(upper curve) and black holes (lower curve), Model BH8.}
   \label{fig:mt}
\end{figure}

A comparison between neutron stars and BHs for the last model is shown in Fig.~3,
where the mean distance for the neutron stars is also evaluated using the
geometric mean.
We note the steady trend for the final 41 neutron stars to segregate towards
the centre, with only six remaining BHs of mean mass $4~M_{\odot}$.
Thus on a slightly longer time-scale the core would be dominated by the
neutron stars, nearly all of mass $1.4~M_{\odot}$.
This difficult and long calculation consumed nearly 300 hours but even so, the
accumulated energy error of $7 \times 10^{-5}$ is highly satisfactory for
such a demanding integration.
Finally, the general expansion of the whole cluster amounts to a factor of 3
as measured by the half-mass radius, in agreement with a standard
calculation (Gieles et al. 2010, eq.~6), while the actual binding energy is
reduced by 8.
Some of the early expansion of the massive component is undoubtedly due to
the effects of mass loss from evolving stars.
It may also be noted that the mass segregation and formation of a compact
subsystem due to dynamical friction is in accordance with earlier theoretical
analysis (Spitzer 1969) although a direct comparison would be difficult.

\section{MODELS}

In this paper, we are mainly interested in the behaviour of black holes.
Before looking at detailed results, we summarize some characteristic stages in
a selected model displayed below, where $i$ denotes the inclination.
Here the key word "Subsystem" means that at least two BHs are sufficiently
compact for the new treatment.

\begin{itemize}
\item BH binary, ~$t = 465$, ~$N_{\rm bh} = 9$, ~$a = 7 \times 10^{-5} $
\item Subsystem, ~$t = 446$, ~$a = 3 \times 10^{-5}$, ~$e = 0.75$
\item Shrinkage, ~$t = 523$, ~$a = 1.8 \times 10^{-5}$, ~$e = 0.65$
\item Kozai cycles, ~$t = 552$, ~$i = 98$, ~$T_{\rm Kozai} = 0.1$~Myr
\item Eccentricity, ~$t = 553$, ~$a = 8 \times 10^{-6}$, ~$e = 0.9996$
\item Coalescence, ~$t = 554$, ~$a = 1 \times 10^{-11}$, ~$e = 0.03$
\item Eccentricity, ~$t = 564$, ~$a = 3 \times 10^{-4}$, ~$e = 0.9999$
\item Shrinkage, ~$t = 638$, ~$a = 3 \times 10^{-5}$, ~$e = 0.67$
\item Subsystem, ~$t = 718$, ~$a = 2 \times 10^{-5}$, ~$e = 0.999$
\item Coalescence, ~$t = 720$, ~$a = 4 \times 10^{-8}$, ~$\tau_{\rm GR} = 1$
\end{itemize}

A few key features in Model~BH12 are displayed at different times in Myr.
This model is somewhat unusual in that there are two coalescence events, with
the most massive BH formed by successive mergers, first by combining 11.9 and
$13.4~M_{\odot}$, followed by the accretion of $15.6~M_{\odot}$.
In spite of continuing the calculation another 700~Myr, the massive BH remained
in the system because during this stage the total mass of the other BH members
only amounted to $19~M_{\odot}$.
We note that each coalescence was preceded by high eccentricity which resulted
in time-consuming post-Newtonian calculations until the condition~(\ref{eq:Rcoal})
was reached or the time-scale became sufficiently short.

\begin{figure}
 \includegraphics[width=8.cm]{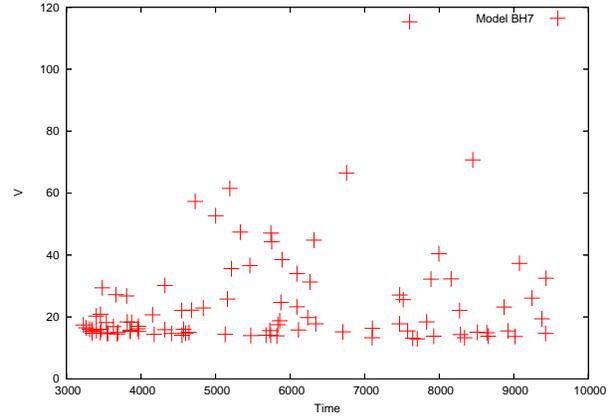} 
     \caption{~~Ejection velocities in ${\rm km~s^{-1}}$ from the central BH
                binary, Model BH7.}
   \label{fig:mt}
\end{figure}

The central concentration of BHs eventually leads to strong interactions.
Particularly energetic ejections occur after the formation of a dominant binary
and at some stage even the binary may escape due to the recoil effect.
All stars with high ejection velocities in Model~BH7 are shown in Fig.~4.
Here the velocity is evaluated with respect to the subsystem c.m.,
usually a binary, taking the lower limit of twice the current average velocity
which typically ensures escape.
In this model, the binary itself was ejected by recoil at time 9600 with velocity
$v_{\rm ej} = 20~{\rm km~s^{-1}}$ and actual velocity at escape
$v_{\rm esc} = 10~{\rm km~s^{-1}}$.
There were four other BHs with terminal escape velocity above $20~{\rm km~s^{-1}}$
but no dynamically ejected neutron stars above half this value\footnote{The
requirement for parabolic escape is
$v_{\rm esc} \simeq 2~{\rm km~s^{-1}} $ at $2 r_{\rm tide}$.}.
In spite of these strong interactions, the numerical accuracy was maintained well
with the accumulated energy error only amounting to $2 \times 10^{-5}$.
It is also worth noting that the average BH mass declines to about $6~M_{\odot}$
which is a direct consequence of mass segregation where the most massive BHs are
ejected first.

\begin{figure}
 \includegraphics[width=8.cm]{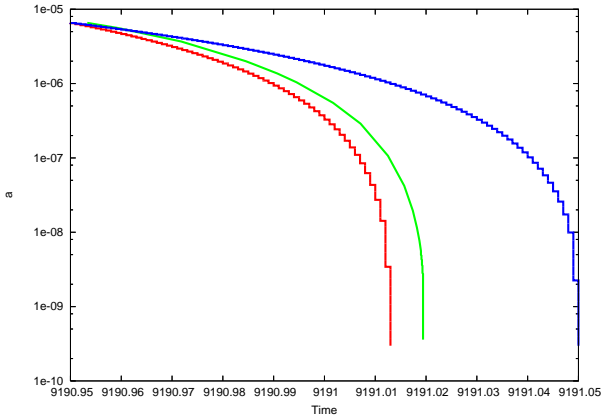} 
     \caption{~~Semi-major axis as function of $N$-body time.
                Middle curve: actual plot from Model~BH12.
                Upper and lower curve: orbit-averaged solution from Peters (1964)
                for two initial values of eccentricity, 0.9996 and 0.99965,
                where only the four first significant figures are available.
                The plot is truncated at $3 \times 10^{-10}$ while the calculation
                extends to $1.5 \times 10^{-11}$. }
   \label{fig:mt}
\end{figure}

Although energetic, the ejection of massive binaries invariably occurs using the
Newtonian formulation where the semi-major axis is still within a modest factor
of $a_{\rm hard}$.
Consequently, favourable configurations are initiated through the Kozai mechanism
where eccentricity growth is possible for long-lived triple systems.
If a sufficiently large value is reached, the semi-major axis and eccentricity
starts decaying according to well-known relativistic expressions (Peters 1964).
The subsequent evolution can take several forms, depending on the influence of
any perturbers.

Figure~5 (middle curve) shows the decreasing semi-major axis due to the full
post-Newtonian treatment until coalescence at $a_{\rm coal} \simeq 1.5 \times 10^{-11}$
or $6 \times 10^{-4}~R_{\odot}$.
Note the short time interval corresponding to $\simeq 4 \times 10^3$~yr required
for the major part of this inspiral.
Also plotted are the orbit-averaged solutions from Peters (1964).
Because the initial eccentricity is only available to four significant figures, we
have included a trial solution for a possible maximum value (lower curve) which
brackets the actual solution.
In view of the uncertainty, the qualitative agreement is satisfactory.
The late stages of shrinkage are characterized by negligible dynamical perturbations.
Nevertheless, in the absence of special procedures, a large number of orbits needs
to be integrated.

\begin{figure}
 \includegraphics[width=8.cm]{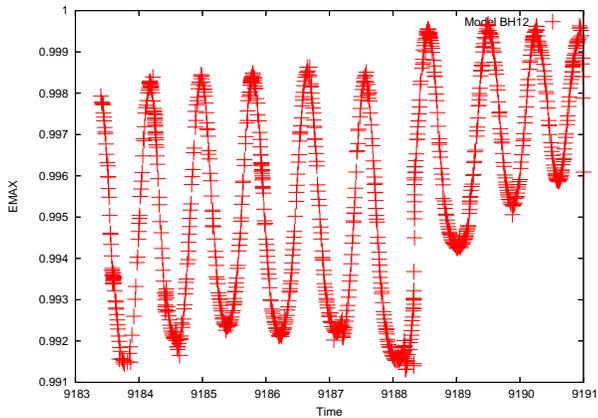} 
     \caption{~~Maximum eccentricity as function of time, Model BH12.
                Note the late increase of eccentricity due to Kozai cycles. }
   \label{fig:mt}
\end{figure}

The corresponding plot is shown in Fig.~6 of the maximum eccentricity
(Heggie 1995, personal communication).
This has been converted to mainly relativistic form\footnote{With two terms of $e^2$
replaced by relativistic values.} and illustrates the role of the Kozai mechanism.
In fact, a large value exceeding 0.998 was already reached some 0.4~Myr earlier but
the most significant shrinkage took place after the final maximum as well as the
actual eccentricity exceeded 0.999.
Since the eccentricity did not show any evidence of decrease during the interval,
this behaviour is an example of so-called Kozai migration (Wu \& Murray 2003).
However, the earlier eccentricity maximum would be sufficient to ensure gravitational
coalescence on a slightly longer time-scale, provided the favourable configuration
is preserved.
It can also be seen that the Einstein shift given by equation~(\ref{eq:dw}) would still
be small during this episode, consistent with the derived value $T_{\rm Kozai} \simeq 1$.

Including all the PN terms up to 3PN demonstrates that the coalescence condition
of equation~(\ref{eq:Rcoal}) can be reached at the maximum eccentricity in agreement
with estimates of the time-scale~(\ref{eq:TGR}).
This is by no means assured since it is known that the relativistic precession acts
to de-tune eccentricity growth.
In this connection, a discussion of the post-Newtonian modification of the Hamiltonian
is of interest (Miller \& Hamilton 2002).
It turns out that the modified expression for the maximum eccentricity still admits
values above 0.999 for the parameters of interest here (Aarseth 2007).

\begin{table}
\caption{Summary of coalescence events.
The coalescence time in Myr is given in Column 2, followed by the corresponding
semi-major axis, eccentricity and combined mass in $M_{\odot}$.}
\begin{tabular}{lllll}
\hline
Model & $T_{\rm coal}$ & $a$ & $e$ & $m$ \\
\hline
BH8   & 300  & $2 \times 10^{-11}$ & 0.00   & 35  \\
BH11  & 184  & $2 \times 10^{-11}$ & 0.9999 & 30  \\
BH12  & 554  & $1 \times 10^{-11}$ & 0.00   & 25  \\
BH12  & 720  & $4 \times 10^{-08}$ & 0.47   & 41  \\
BH13  & 716  & $3 \times 10^{-07}$ & 0.99   & 28  \\
BH14  & 360  & $6 \times 10^{-07}$ & 0.985  & 29  \\
BH14  & 760  & $2 \times 10^{-06}$ & 0.9997 & 14  \\
\hline
\end{tabular}
\end{table}

A summary of all the coalescence events is given in Table~1.
A total of 14 models have been investigated using the PN formulation.
However, a few were terminated prematurely because of technical difficulties.
Most of the other models were continued further, unless the total BH mass was
small.
Although all the cases listed were accepted as coalescence, some did not reach
the actual end state of equation~(\ref{eq:Rcoal}) because one of the secondary
criteria discussed above was satisfied.
For example, in the case of Model~BH11, the eccentricity was extremely
large for a short time interval.
Likewise, for Model~BH14 the high eccentricity and short time-scale would
ensure coalescence.

\begin{figure}
 \includegraphics[width=8.cm]{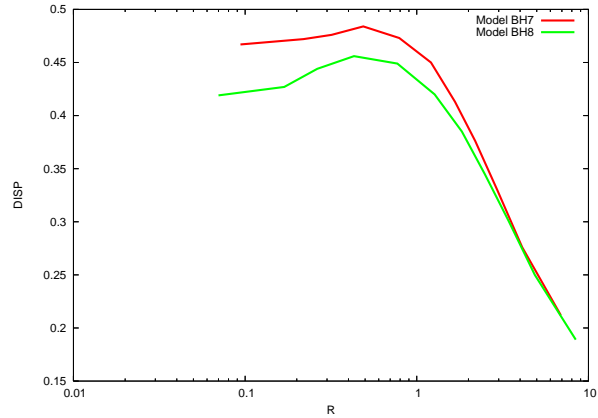} 
     \caption{~~Final velocity dispersion profiles as a function of central distance,
                Model BH7 (upper curve) and BH8 (lower curve).
                The current density centre is used as reference point
                and the respective times are 750~Myr and 1.1~Gyr. 
                BHs and neutron stars have been excluded but the small membership
                hardly affects the outcome. }
   \label{fig:mt}
\end{figure}

 There have been many observational efforts to determine the velocity
 dispersion as a function of radius, in particular related to the quest for
 discovering an intermediate mass BH in a globular cluster.
 The velocity profiles of all luminous members in two models are shown in Fig.~7 for
 illustration.
 Here the more evolved Model~BH8 exhibits a larger central velocity but both models
 are characterized by a slight velocity decrease in the innermost region.
 This deficiency may be expected to disappear as the cluster is depleted of BHs.
 We note that in an investigation involving a population of BHs, the central
 velocity profile (line-of-sight) showed a flat slope inside 1~pc
(Mackey et al.~2008, Fig.~12).
 However, a direct comparison is difficult because here some 200 BHs were retained.

\begin{figure}
 \includegraphics[width=8.cm]{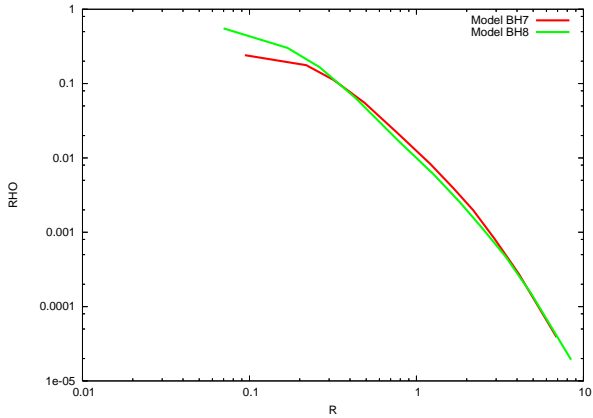} 
     \caption{~~Final density profiles, Model BH7 (upper curve) at 750~Myr and
                BH8 (lower curve) at 1.1~Gyr.
                A small number of BHs and neutron stars are included. }
   \label{fig:mt}
\end{figure}

The density is also of considerable interest.
The final densities of the two models shown in Fig.~8 are quite well matched.
Consistent with the small velocity decrease towards the centre, the central density
profile has a flatter slope.
Again this feature will tend to disappear as the post-BH core formation proceeds.

\begin{figure}
 \includegraphics[width=8.cm]{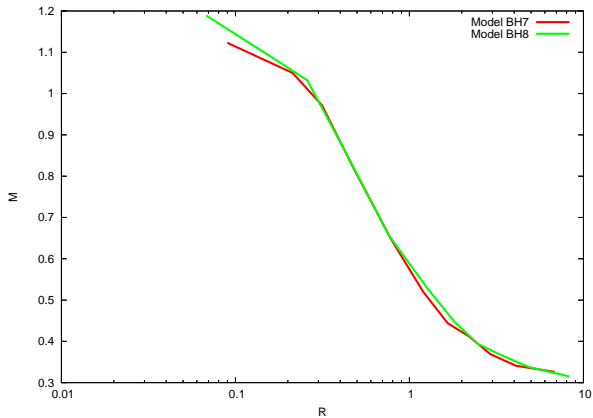} 
     \caption{~~Average stellar mass in solar units as a function of central distance.
                The plots for Model BH7 at 750~Myr and BH8 at 1.1~Gyr are very
                similar, with the upper stellar mass limits near $3~M_{\odot}$.}
   \label{fig:mt}
\end{figure}

The process of mass segregation plays an important role in the present work.
In Fig.~9 we illustrate the average mass as a function of central distance in two
typical models where BHs and neutron stars have been excluded.
At the end states the maximum mass has been reduced to $3~M_{\odot}$
due to stellar evolution while the minimum mass used is $0.1~M_{\odot}$.
Although the final times differ somewhat (750~Myr vs 1.1~Gyr), the two  plots
are essentially indistinguishable and span a significant mass range.

\section{DISCUSSION}

The main purpose of this paper is to demonstrate the numerical capability of
the code {\tt NBODY7} for studying strong interactions in cluster cores.
It turns out that the algorithmic regularization method is also able to treat
post-Newtonian terms with high accuracy.
In fact, energy conservation is now substantially better than in the standard
{\tt NBODY6} code because the dominant errors are associated with the most
strongly bound members and the {\tt ARC} method is more accurate.
This improved treatment comes at a price but already it is feasible to investigate
small globular clusters using realistic parameters.

The {\tt ARC} software usually maintains high accuracy as a result of specifying a
small tolerance in the Bulirsch--Stoer integrator combined with a more efficient
time transformation.
A further advantage compared to standard NBODY6 integration with chain
regularization is that the subsystems are treated for much longer times so there
are less errors due to switching.
In any case, a direct comparison with the chain code is only possible in the
absence of PN terms.
Moreover, the latest version of {\tt ARC} admits solutions with only two members
when PN terms are present.
It is also helpful that the smaller density in the inner region reduces the
energy errors associated with the single particles.

The onset of conditions for gravitational radiation is usually initiated by the
Kozai mechanism.
Following the subsequent shrinkage, the inner binary orbit may become detached
(or isolated) so that the final stages are often unperturbed dynamically.
This behaviour would enable a sequential analytical continuation of this
time-consuming process, taking into account the Einstein shift of
equation~(\ref{eq:dw}) or the addition of its second-order equivalent\footnote
{The analytical continuation procedure has been implemented recently.}.

Stellar systems containing a population of massive objects experience mass
segregation on a short time-scale.
Although the velocity kick at formation increases the velocity dispersion of
neutron stars and BHs, the remaining BHs soon begin to concentrate towards the
centre.
However, the emergence of a dominant binary acts to deplete the central density
and prevents the formation of other binaries, an aspect which has not been
considered so far.

The problem of black hole dynamics has been studied by many authors.
Most studies, however, have been concerned with choices of the IMF which are
more applicable to intermediate-mass BHs and hence comparisons are not
appropriate.
The requirement of realistic PN treatment is also rarely addressed.
For these reasons, we only mention one recent work based on stellar mass BHs
(Banerjee, Baumgardt \& Kroupa 2010).
Here an IMF truncated to low-mass stars ($m < 1~M_{\odot}$) was adopted with
an addition of about 100 BHs of $10~M_{\odot}$.
Although not a fully consistent simulation, the characteristic behaviour of
mass segregation was seen, together with ejection of massive members.
Moreover, employment of the relativistic orbit-averaged changes in semi-major
axis and eccentricity (Peters 1964) gave rise to several cases of coalescence.

Contrary to the evidence for BH binary formation, a full simulation by {\tt NBODY7}
with all 140 BHs retained showed that few binaries are formed dynamically during
the first Gyr when half the BHs have escaped.
This behaviour can be understood in terms of an emerging binary being exposed to
disruptive encounters with the dominant binary before becoming hard.
There remains the possibility, however, that a dominant binary may be ejected
temporarily from the core by recoil, giving the more weakly binary an opportunity
of hardening.

The question of BH coalescence by the Kozai mechanism in long-lived triple
systems has also been discussed elsewhere (Miller \& Hamilton 2002).
However, this work assumes a certain fraction of BH binaries in order for
binary--binary collisions to take place and is therefore not consistent with
the current findings.
Given the relatively small number of BHs expected from a realistic IMF, the
potential for massive binary formation is fairly limited.
Moreover, the possible depletion due to natal velocity kicks should also be
taken into account.

For completeness, a runaway scenario for compact objects was also studied with a
full post-Newtonian formulation (Kupi, Amaro-Seoane \& Spurzem 2006).
In this early PN application with the code {\tt NBODY6++}, the treatment of all
close two-body encounters was modified by including the effect of relativistic
perturbations.
Given the large central velocity dispersion of $4300~{\rm km~s^{-1}}$, a
massive object was formed by runaway accretion after core collapse.
We note that the Kustaanheimo--Stiefel (1965) Hermite implementation also
requires the first time derivative of the relevant PN force terms while in
algorithmic or chain regularization only the basic expressions of
equation~(\ref{eq:GR}) are needed.

\section{CONCLUSIONS}

This is a first $N$-body implementation of the new algorithmic regularization
scheme (Hellstr\"om \& Mikkola 2010).
The application to a system containing black holes poses certain numerical
problems, mainly in connection with membership selection.
Although the post-Newtonian terms require time-consuming calculations, it has
proved possible to describe the full binary evolution up to coalescence using
realistic parameters.
Further progress can be made taking advantage of analytical continuation of
unperturbed binaries in the relativistic regime after allowing for the precession.
The question of how to initialize the spin must also be addressed.
Its contribution to the post-Newtonian terms enters at the order $c^{-3}$ and
is therefore more important than the 2PN terms\footnote{Spin and a small merger
recoil velocity have now been implemented.}.
On the observational side, we emphasize the presence of additional escaping stars
of high velocity, especially in the later stages when massive stellar binaries
are no longer present.

The present results indicate that there is no evidence for runaway evolution of
BH masses.
Given the relatively small number of BHs involved, just one binary may dominate
the central region.
Occasionally strong interactions lead to fast ejection of other BHs and even the
binary may escape in a slingshot event.
These simulations show that even a small number of BHs may play an important
role in the first Gyr of globular cluster evolution.
Since the simulations only cover the first Gyr so far, later stages may repopulate
the core in accordance with expectations.
In the absence of a definite model for natal BH velocity kicks, further
explorations of larger populations are desirable.
We have seen that the Kozai mechanism can be very effective in leading to the
relativistic coalescence of BH binaries.
As far as stellar mass BHs are concerned, sufficiently high eccentricities can
be reached without the growth being de-tuned by precession effects.
Finally, the presence of even a few stellar mass black holes with luminous
companions may provide a new challenge for future observations.

\section{ACKNOWLEDGMENTS}

Vital contributions to this work were made by Keigo Nitadori who implemented
the GPU system and Seppo Mikkola who provided the software for algorithmic
regularization.
I am also indebted to the referee for many constructive suggestions which
helped to improve the manuscript.


\begin{thebibliography}{99}

\bibitem[\protect\citeauthoryear{Aarseth}{1971}]{aa71}
Aarseth S.~J., 1971, Ap\&SS, 13, 324
%massive binary evolution 

\bibitem[\protect\citeauthoryear{Aarseth}{2003}]{aa03}
Aarseth S.~J., 2003a, Gravitational N-Body Simulations, Cambridge
Univ. Press, Cambridge

\bibitem[\protect\citeauthoryear{Aarseth}{2003b}]{aa03b}
Aarseth S.~J., 2003b, Ap\&SS, 285, 367
%first demonstration of full PN and large ECC

\bibitem[\protect\citeauthoryear{Aarseth}{2007}]{aa07}
Aarseth S.~J., 2007, MNRAS, 378, 285
%PN

\bibitem[\protect\citeauthoryear{Aarseth \& Zare}{1974}]{az74}
Aarseth S.~J., Zare K., 1974, Celes.~Mech.,~10, 185
%AZ

\bibitem[\protect\citeauthoryear{Banerjee, Baumgardt \& Kroupa}{1973}]{bb09}
Banerjee S., Baumgardt H., Kroupa P., 2010, MNRAS, 402, 371
%black holes in star clusters, mass segregation of BHs

\bibitem[\protect\citeauthoryear{Berentzen \& Preto}{2009}]{bp09}
Berentzen I., Preto M., Berczik P., Merritt D., Spurzem R., 2009, ApJ,~695, 455
%m1=m2=0.01, incomplete PN evolution at termination

\bibitem[\protect\citeauthoryear{Blanchet \& Iyer}{2003}]{bi03}
Blanchet L., Iyer B., 2003, Class.~Quant.~Grav.,~20, 755

\bibitem[\protect\citeauthoryear{Brockamp, Baumgardt \& Kroupa}{2011}]{bb11}
Brockamp M., Baumgardt H., Kroupa P., 2011, MNRAS, 418, 1308
%tidal disruption rate by SMBH without PN

\bibitem[\protect\citeauthoryear{Bulirsch \& Stoer}{1966}]{bs66}
Bulirsch R., Stoer J., 1966, Num.~Math.,~8, 1

\bibitem[\protect\citeauthoryear{Einstein}{1915}]{e15}
Einstein A., 1915, Sitzungsber.~Preuss.~Acad.~Wiss., 47, 2, 831

\bibitem[\protect\citeauthoryear{Eldridge \& Tout}{2004}]{et04}
Eldridge J., Tout C.~A., 2004, MNRAS, 353, 87E

\bibitem[\protect\citeauthoryear{Gieles \& Baumgardt}{2004}]{gb10}
Gieles M., Baumgardt H., Heggie D.~C., Lamers J.~G.~L.~M., 2010, MNRAS, 408, 16

\bibitem[\protect\citeauthoryear{Harfst \& Gualandris}{2008}]{hg08}
Harfst S, Gualandris A., Merritt D., Mikkola S., 2008, MNRAS, 389, 2
%hybrid code with AR-chain

\bibitem[\protect\citeauthoryear{Hellstr\"om \& Mikkola}{2010}]{hm10}
Hellstr\"om C., Mikkola S., 2010, Celes.~Mech.~Dyn.~Ast., 106, 143

\bibitem[\protect\citeauthoryear{Hurley, Pols \& Tout}{2000}]{hp00}
Hurley J.~R., Pols O., Tout C.~A., 2000, MNRAS, 315, 543
%SSE package

\bibitem[\protect\citeauthoryear{Hurley, Pols \& Tout}{2002}]{hp02}
Hurley J.~R., Pols O., Tout C.~A., 2002, MNRAS, 329, 897
%BSE package

\bibitem[\protect\citeauthoryear{Iwasawa, Funato \& Makino}{2006}]{if06}
Iwasawa M., Funato Y., Makino J., 2006, ApJ, 651, 1059
%massive triple BHs with GRAPE6, small c

\bibitem[\protect\citeauthoryear{Kiseleva, Eggleton \& Mikkola}{1998}]{ke98}
Kiseleva L.~G., Eggleton P.~P., Mikkola S., 1998, MNRAS, 300, 292
%Kozai period

\bibitem[\protect\citeauthoryear{Kozai}{1962}]{ko62}
Kozai Y., 1962, AJ,~67, 591
%original Kozai cycles paper

\bibitem[\protect\citeauthoryear{Kozai}{2004}]{ko04}
Kozai Y., 2004, Proc.~Jpn.~Acad., Ser.~B, 80, 157
%theory extension

\bibitem[\protect\citeauthoryear{Kroupa, Tout \& Gilmore}{1993}]{kt93}
Kroupa P., Tout C., Gilmore G., 1993, MNRAS, 262, 545
%IMF extended to BD region

\bibitem[\protect\citeauthoryear{Kupi, Amaro-Seoane \& Spurzem}{2006}]{ka06}
Kupi G., Amaro-Seoane P., Spurzem R., 2006, MNRAS, 371, L45
%fully relativistic cluster with up to 2.5PN in NBODY6; runaway BH

\bibitem[\protect\citeauthoryear{Kustaanheimo \& Stiefel}{1965}]{ks65}
Kustaanheimo P., Stiefel E., 1965, J.~Reine Angew.~Math. ~218, 204
%KS

\bibitem[\protect\citeauthoryear{Mackey et al.}{2008}]{mw08}
Mackey A.~D., Wilkinson M.~L., Davies M.~B., Gilmore G.~F., 2008,
MNRAS, 386, 65
%BH simulation on GRAPE-6

\bibitem[\protect\citeauthoryear{Mikkola \& Aarseth}{1993}]{ma93}
Mikkola S., Aarseth S.~J., 1993, Celes.~Mech.~Dyn.~Ast., 57, 439
%chain regularization

\bibitem[\protect\citeauthoryear{Mikkola \& Aarseth}{2002}]{ma02}
Mikkola S., Aarseth S.~J., 2002, Celes.~Mech.~Dyn.~Ast., 84, 343
%TTL

\bibitem[\protect\citeauthoryear{Mikkola \& Merritt}{2006}]{mm06}
Mikkola S., Merritt D., 2006, MNRAS, 372, 219
%AR time transformation

\bibitem[\protect\citeauthoryear{Mikkola \& Merritt}{2008}]{mm08}
Mikkola S., Merritt D., 2008, A.~J. 135, 2398

\bibitem[\protect\citeauthoryear{Mikkola \& Tanikawa}{1999}]{mm99}
Mikkola S., Tanikawa K., 1999, MNRAS, 310, 745

\bibitem[\protect\citeauthoryear{Miller \& Hamilton}{2002}]{mh02}
Miller M.~C., Hamilton D.~P., 2002, ApJ, 576, 894

\bibitem[\protect\citeauthoryear{Mora \& Will}{2004}]{mw04}
Mora T., Will C., 2004, Phys.~Rev.~D 69, 104021

\bibitem[\protect\citeauthoryear{Nitadori}{2009}]{n09}
Nitadori K., 2009, Ph.~D. ~thesis, University of Tokyo

\bibitem[\protect\citeauthoryear{Nitadori \& Aarseth}{2012}]{na12}
Nitadori K., Aarseth S.~J., 2012, MNRAS, in press

\bibitem[\protect\citeauthoryear{Peters}{1964}]{pe64}
Peters P.~C., 1964, Phys.~Rev., 136, B1224

\bibitem[\protect\citeauthoryear{Spitzer}{1969}]{sp69}
Spitzer L., 1969, ApJ, 158, L139
%mass segregation and compact system formation

\bibitem[\protect\citeauthoryear{van Albada}{1968}]{va68}
van Albada T.~S., 1968, Bull.~Astron.~Inst.~Neth., 19, 479

\bibitem[\protect\citeauthoryear{Wu \& Murray}{2003}]{wm03}
Wu Y., Murray N., 2003, ApJ, 589, 605

\bibitem[\protect\citeauthoryear{Zare}{1974}]{za74}
Zare, K., 1974, Celes.~Mech., 10, 107

\end{thebibliography}
\end{document}